\documentclass[lettersize,journal]{IEEEtran}
\usepackage[T1]{fontenc}

\usepackage[none]{hyphenat}
\usepackage[utf8]{inputenc}
\usepackage[english]{babel}

\usepackage{verbatim}
\usepackage{graphicx}
\usepackage{float}
\usepackage{hyperref}
\usepackage{lipsum}
\usepackage{algorithmic}
\usepackage{algorithm}
\usepackage{textcomp}
\usepackage{stfloats}
\usepackage{url}
\usepackage{amsmath,amsthm}
\usepackage{amssymb}
\usepackage{mathtools}
\usepackage{epstopdf}
\usepackage{booktabs,dcolumn}
\usepackage{multirow}
\usepackage{caption}
\usepackage{subcaption}
\usepackage{ragged2e}
\usepackage{cite}
\usepackage{censor}

\usepackage{array}
\newcolumntype{C}[1]{>{\centering\arraybackslash}p{#1}}
\newcolumntype{L}[1]{>{\RaggedRight\arraybackslash}p{#1}}
\usepackage{wasysym}

\newcommand*\diff{\mathop{}\!\mathrm{d}}

\usepackage{xcolor}
\usepackage{colortbl}

\usepackage{caption,hypcap}

\usepackage{cleveref}
\usepackage{scalerel}
\usepackage{tikz}
\usetikzlibrary{svg.path, decorations.markings}

\usepackage{physics}
\usepackage{ifthen}
\usepackage[outline]{contour} 
\usetikzlibrary{calc} 
\usetikzlibrary{angles,quotes} 
\usetikzlibrary{patterns}

\usepackage{enumitem}

\usepackage{soul}

\definecolor{orcidlogocol}{HTML}{A6CE39}
\tikzset{
  orcidlogo/.pic={
    \fill[orcidlogocol] svg{M256,128c0,70.7-57.3,128-128,128C57.3,256,0,198.7,0,128C0,57.3,57.3,0,128,0C198.7,0,256,57.3,256,128z};
    \fill[white] svg{M86.3,186.2H70.9V79.1h15.4v48.4V186.2z}
                 svg{M108.9,79.1h41.6c39.6,0,57,28.3,57,53.6c0,27.5-21.5,53.6-56.8,53.6h-41.8V79.1z M124.3,172.4h24.5c34.9,0,42.9-26.5,42.9-39.7c0-21.5-13.7-39.7-43.7-39.7h-23.7V172.4z}
                 svg{M88.7,56.8c0,5.5-4.5,10.1-10.1,10.1c-5.6,0-10.1-4.6-10.1-10.1c0-5.6,4.5-10.1,10.1-10.1C84.2,46.7,88.7,51.3,88.7,56.8z};
  }
}

\newcommand\orcidicon[1]{\href{https://orcid.org/#1}{\mbox{\scalerel*{
\begin{tikzpicture}[yscale=-1,transform shape]
\pic{orcidlogo};
\end{tikzpicture}
}{|}}}}

\usepackage{suffix}

\usepackage{amsmath,amssymb,amsfonts}
\usepackage{algorithmic}
\usepackage{graphicx,color}
\usepackage{textcomp}

\usepackage[none]{hyphenat}
\usepackage[utf8]{inputenc}
\usepackage[english]{babel}

\usepackage{verbatim}
\usepackage{graphicx}
\usepackage{subcaption}
\usepackage{float}
\usepackage{booktabs}
\usepackage{mathtools}
\usepackage{hyperref}
\usepackage{lipsum}
\usepackage{stfloats}
\usepackage{amsmath,amsthm}
\usepackage{amssymb}
\usepackage{mathtools}
\usepackage{epstopdf}
\usepackage{booktabs,dcolumn,caption}
\usepackage{multirow}
\usepackage{caption}
\usepackage{subcaption}
\usepackage{ragged2e}
\usepackage{censor}
\usepackage{tabularx}
\usepackage{relsize}
\usepackage{pgfplots}
\usepackage{url}
\usepackage{textcomp}
\usepackage{parskip}
\usepackage[numbers]{natbib}
\usepackage{graphicx}

\usepackage{booktabs} 
\usepackage{siunitx}  

\usepackage{array}
\newcolumntype{C}[1]{>{\centering\arraybackslash}p{#1}}
\newcolumntype{L}[1]{>{\RaggedRight\arraybackslash}p{#1}}
\usepackage{wasysym}
\usepackage{mathtools}
\newcommand\Set[2]{\{\,#1\mid#2\,\}}
\newcommand\SET[2]{\Set{#1}{\text{#2}}}

\usepackage{mathrsfs}
\usepackage{colortbl}
\usepackage{color,soul}

\usepackage{caption,hypcap}

\usepackage{cleveref}

\usepackage{suffix}
\usepackage{mathtools, cuted}
\usepackage{lipsum, color}

\usepackage{enumitem}

\usepackage{pgfplots}
\usepackage{tikz}

\usetikzlibrary{angles,quotes}

\usepackage{mathtools, cuted}

\DeclarePairedDelimiter{\ceil}{\lceil}{\rceil}

\DeclarePairedDelimiter{\floor}{\lfloor}{\rfloor}

\pgfmathdeclarefunction{invgaussx}{4}{%
    \pgfmathparse{#1+#2*sqrt(-2*ln(#3))*cos(deg(2*pi*#4))}%
}
\pgfmathdeclarefunction{invgaussy}{4}{%
\pgfmathparse{#1+#2*sqrt(-2*ln(#3))*sin(deg(2*pi*#4))}%
}

\newcommand{\itemEq}[1]{%
        \begingroup%
        \setlength{\abovedisplayskip}{-5pt}%
        \setlength{\belowdisplayskip}{0pt}%
        \parbox[c]{\linewidth}{\begin{flalign}#1&&\end{flalign}}%
        \endgroup}

\usepackage{amssymb}
\usepackage{amsmath}

\hypersetup{linkcolor=black,citecolor=black}

\begin{document}

\title{On the Conditional Phase Distribution of the TWDP Multipath Fading Process}

\author{Almir~Maric\textsuperscript{\orcidicon{0000-0001-5912-2967}}~\IEEEmembership{}
        and~Pamela~Njemcevic\textsuperscript{\orcidicon{0000-0002-3005-3934}}~\IEEEmembership{}
\thanks{This research is funded by Federal Ministry of Education and Science, Bosnia and Herzegovina (grant number).}
\thanks{A. Maric and P. Njemcevic are with Department of Telecommunications, Faculty of Electrical Engineering, University of Sarajevo, Sarajevo, B\&H.}
\thanks{e-mail:\{almir.maric,pamela.njemcevic\}@etf.unsa.ba}
}


\IEEEpubid{\begin{tabular}[t]{@{}c@{}}This work has been submitted for possible publication.\\Copyright may be transferred without notice, after which this version may no longer be accessible.\end{tabular}}

\maketitle

\begin{abstract}
\label{sec:abs}
In this paper, the conditional phase distribution of the two-wave with diffuse power (TWDP) process is derived as a closed-form and as an infinite-series expression. For the obtained infinite series expression, a truncation analysis is performed and the truncated expression is used to examine the influence of different channel conditions on the behavior of the TWDP phase. All the results are verified through Monte Carlo simulations.
\end{abstract}


\section{Introduction}
\label{sec:I}
\IEEEPARstart{T}{wo}-wave with diffuse power (TWDP) is a multipath fading model introduced to model better-than-Rayleigh and worse-than-Rayleigh fading conditions~\cite{Dur02}. The model has been empirically verified in vehicular-to-infrastructure (V2I), vehicle-to-vehicle (V2V) and indoor propagation environments at 60 GHz, as well as in wireless sensor networks deployed in cavity environments~\cite{Zoc19-1, Zoc19-2, Zoc19, Fro08}. It also incorporates Rayleigh and Rician models as special cases, which  makes it applicable for modeling multipath fading in various propagation scenarios encountered in the emerging wireless networks.

Given its significance, TWDP envelope statistics have been studied extensively~\cite{Dur02, Rao15, rad}. However, characteristics of the received signal envelope reveal only part of the overall channel picture~\cite{Bro23}. Therefore, to fully characterize the fading process, the statistical properties of the corresponding phase are also of interest, since they can be used in the design and synchronization of coherent receivers and in the detection of M-ary phase shift keying signal
constellations~\cite{Bro23}. More precisely, the knowledge about the phase behavior can be utilized for the optimal design and synchronization of coherent receivers,
 where carrier recovery schemes are required~\cite{Por16}. Additionally, the performance of modulation systems using non-ideal coherent or incoherent detection, as well as those employing an orthogonal frequency-division multiple access (OFDM) modulation scheme, is also affected by the phase statistics of the communication channel~\cite{Por16}. Accordingly, understanding
the properties of the phase distribution is essential for the design of wireless systems~\cite{Bro19}.

However, despite its importance, and the fact that the probability distribution function (PDF) of the phase process has been derived and discussed for many multipath fading models (such as Rician~\cite{Tih75}, Nakagami-m~\cite{Por13, Yac10}, $\eta-\mu$~\cite{Ben07}, $\kappa-\mu$~\cite{Por16}, etc.), no results regarding the phase behavior have been provided for the TWDP model. The only relevant results are published in~\cite{Kos82}, where the phase probability density function of the sum of the signal, noise and cochannel interference is considered under equivalent statistical assumptions to those of the TWDP model~\cite{rad}.

In this paper, we derived the conditional PDF of the TWDP phase stemming from the fundamental underlying assumptions of the TWDP model itself, and then, investigated the impact of different multipath fading conditions on the behavior of the phase process.  
Accordingly, after the Introduction given in Section~\ref{sec:I}, 
the conditional TWDP phase PDF is foremost derived as an infinite-series expression in Section~\ref{sec:II}. This is followed by an appropriate truncation analysis, given in Section~\ref{sec:IV}, and derivation of the closed-form expression in Section~\ref{sec:III}. Numerical results, presented in Section~\ref{sec:V}, verified the obtained analytical results, providing for the first time in the literature, insight into the behavior of the TWDP phase process and its specific features. The derived analytical expressions are also verified using the developed TWDP geometrically-based simulator in Section~\ref{sec:VI}, and the results are finally applied to error performance evaluation in Section~\ref{sec:VII}. Final conclusions and guidelines for future work are given in Section~\ref{sec:VII}.

\section{Conditional TWDP Phase PDF}
\label{sec:II}

TWDP is a multipath fading model applicable for modeling various fading conditions. It assumes that the complex envelope $\rho(t)$ of the signal propagating in the TWDP channel 
is composed of two strong specular components $v_1(t)$ and $v_2(t)$, along with the diffuse component $n(t)$~\cite{Dur02}:
\begin{equation}
\label{definicija}
\begin{split}
    \rho(t) & =  v_1(t) + v_2(t) + n(t) \\ & = V_1\exp{\left(j\varPhi_1\right)} + V_2\exp{\left(j\varPhi\right)} + n(t)
    \end{split}
\end{equation} 
\IEEEpubidadjcol It also assumes that the specular components have constant magnitudes $V_1$, $V_2$, where $V_2 \leq V_1$, and uniformly distributed phases $\varPhi_1$, $\varPhi$ in the range $[-\pi,\pi]$, while the diffuse component is considered as a zero mean Gaussian random process with average power $2\sigma^2$.

\begin{figure}[t]
	\centering
    \vspace{-0.25cm}
    \tikz \node [scale=0.99] {
    \begin{tikzpicture}[>= latex]
        \def\VIT{5};
        \def\VIIT{2.5};
        \def\PhiT{45.1};
        \def\SigT{0.25};
        \def\XT{\VIIT*cos(\PhiT)-\SigT};
        \def\YT{\VIIT*sin(\PhiT)+2*\SigT};
        \def\ThetaT{atan2(\YT,\XT)};
        \begin{axis}[
                set layers,mark layer=axis background,
                grid=major,
                axis equal image,
                xmin=-5.5,xmax=2.5,samples=201,
                xlabel=$x$,ylabel=$y$,
                ymin=-0.7,ymax=2.7,
                restrict y to domain=-1:3,
                enlargelimits={abs=0.1cm},
                axis line style={latex-latex},
                axis x line=center,
                axis y line=center,
                xtick={\empty},ytick={\empty},
                xlabel style={at={(ticklabel* cs:1)},anchor=west,font=\scriptsize},
                ylabel style={at={(ticklabel* cs:1)},anchor=north,font=\scriptsize},
                mark options={mark=*, mark size=0.6pt, gray, fill=gray},
            ]
            \addplot [thick,black,->] coordinates {(-\VIT,0) (0,0)} node[below left,pos=1,font=\normalsize] {$V_1$};
            \addplot table [x expr={\VIIT*cos(\PhiT)+\SigT*\thisrow{x}}, y expr={\VIIT*sin(\PhiT)+\SigT*\thisrow{y}}, only marks, mark=o] {points.dat};
            \addplot [thick,black,->] coordinates {(0,0) ({\VIIT*cos(\PhiT)},{\VIIT*sin(\PhiT)})} node[below right,pos=1,font=\normalsize] {$V_2$};
            \addplot [thick,black,->] coordinates {({\VIIT*cos(\PhiT)},{\VIIT*sin(\PhiT)}) (\XT,\YT)} node[right,pos=1.2,font=\normalsize] {$n(t)$};
            \draw [ultra thin,black,->] (axis cs:1,0) arc[start angle=0, end angle=\PhiT, radius={transformdirectionx(1)}] node[left,pos=0.4,font=\normalsize] {$\phi$};
            \addplot [only marks, samples=1, black, mark size=0.5pt, line width=0.3pt, mark options={fill=gray}] ({\XT},{\YT});
            \addplot [thick,red,->] coordinates {(-\VIT,0) (\XT,\YT)} node[above,pos=0.5,font=\normalsize] {$\rho$};
            \addplot [thick,blue,->] coordinates {(0,0) (\XT,\YT)} node[left,pos=0.5,font=\normalsize] {$r$};
            \draw [ultra thin,blue,->] (axis cs:1.5,0) arc[start angle=0, end angle=\ThetaT, radius={transformdirectionx(1.5)}] node[left,pos=0.4,font=\normalsize] {$\vartheta$};
            \draw [ultra thin,red,->] (axis cs:-\VIT+2.0,0) arc[start angle=0, end angle={atan2(\YT,\VIT+\XT)}, radius={transformdirectionx(2.0)}] node[left,pos=0.4,font=\normalsize] {$\varphi$};
        \end{axis}
    \end{tikzpicture}
    };
    \vspace{0.07cm}
	\caption{TWDP phasor diagram illustrating $p(r,\vartheta|\phi)$ for $\varPhi_1=0$}
	\label{Fig_1}
\end{figure}
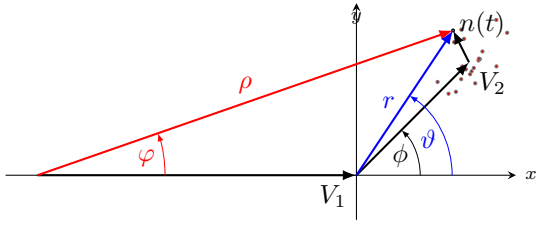

To derive the conditional PDF of the TWDP phase, when the stronger specular component is ideally recovered, without loss of generality can be assumed that $\varPhi_1=0$\footnote{
To simplify the notation, the condition $\varPhi_1=0$ is not explicitly stated in the remaining text.\vspace{-0.8cm}}. In this case, the conditional complex TWDP envelope can be expressed as the sum of $V_1$ and the Rician component $r(t)$, as shown in Fig.~\ref{Fig_1}:
\begin{equation}
\begin{split}
    \rho(t)|_{\varPhi_1=0} = V_1\exp{\left(j 0\right)} +  \underbrace{V_2\exp{\left(j\varPhi\right)} + n(t)}_\text{Rician component, $r(t)$}
    \end{split}
\end{equation} 

Following the aforesaid and the phasor diagram given in Fig.~\ref{Fig_1}, the conditional distribution of the TWDP phase, $p(\varphi)$, can be obtained using one of the following expressions:
\vspace{2mm}
\begin{enumerate}
    \item \itemEq{p(\varphi)=\int_{-\pi}^{\pi} p(\varphi|\phi) p(\phi) d\phi}
    \vspace{2pt}
    \item \itemEq{p(\varphi)=\int_{0}^{\infty} p(\varphi|r) p(r) dr \label{ItemEq}}
\end{enumerate}
The first approach is employed in~\cite{Kostic1979}, where the conditional phase is derived for the sum of signal, cochannel interference and AWGN. It is shown that this phase follows the same statistical characteristics as the components of the TWDP model itself. However, the conditional PDF of the phase derived in~\cite{Kostic1979} is given in the Fourier series form which involves several sums and consequently has a quite complex mathematical notation. 

Accordingly, to obtain more analytically tractable expression, this paper employs the second approach which first requires derivation of the conditional distribution $p(\varphi|r)$, where $r$ is the envelope of the Rician-distributed variable. To achieve this, we started from the joint envelope-phase PDF of the Rician component, given as~\cite{Bro19}: 
\begin{equation}
\begin{split}
\label{uslovna}
    p(r,\vartheta|\phi)=\frac{r}{2\pi\sigma^2} \exp{\left(-\frac{r^2}{2\sigma^2}+\frac{2rV_2}{2\sigma^2}\cos{\left(\vartheta-\phi\right)}-\frac{V_2^2}{2\sigma^2}\right)}
    \end{split}
\end{equation} 
where, by definition, the angle $\phi$ is uniformly distributed 
i.e., ${p(\phi)=1/(2\pi)}$, and $p(\vartheta|\phi)$ is the Rician phase PDF (as can be seen in Fig.~\ref{Fig_1}). Consequently, it can be shown that $\vartheta$ is uniformly distributed between $-\pi$ and $\pi$, i.e., $p(\vartheta)=1/(2\pi)$.

Since $p(r, \vartheta|\phi)$ is invariant with respect to the arbitrary angular shift, i.e., $p(r, \vartheta|\phi) = p(r, \vartheta + \alpha|\phi + \alpha)$, it can be shown that for $\alpha = \phi - \vartheta$, $p(r, \vartheta|\phi)$ is equal to $  p(r, \phi|2\phi - \vartheta)$. Under this condition and after some manipulations of (\ref{uslovna}), it can be concluded that $r$ and $\phi$ are mutually independent, with the corresponding PDFs given as:
\begin{equation}
\begin{split}
\label{Fi}
    p(\phi)=p(\phi|r)=\frac{1}{2\pi}
\end{split}
\end{equation} 
\begin{equation}
\begin{split}
    \label{rajs}
    p(r)=p(r|\phi)&=\frac{r}{\sigma^2} \exp{\left(-\frac{V_2^2+r^2}{2\sigma^2}\right)} I_0\left(\frac{rV_2}{\sigma^2}\right)
\end{split}
\end{equation} 

\begin{figure}[t]
	\centering
    \tikz \node [scale=0.99,below=-0.5cm] {
    \begin{tikzpicture}[>= latex]
        \def\VIT{5};
        \def\VIIT{2.5};
        \def\PhiT{45.1};
        \def\SigT{0.25};
        \def\XT{(\VIIT-3*\SigT)*cos(\PhiT)};
        \def\YT{(\VIIT-3*\SigT)*sin(\PhiT)};
        \def\ThetaT{atan2(\YT,\XT)};
        \pgfmathsetseed{23654}
        \def\XX{invgaussx(0,\SigT,rnd,rnd)};
        \begin{axis}[
                set layers,mark layer=axis background,
                grid=major,
                axis equal image,
                xmin=-6,xmax=4,samples=201,
                xlabel=$x$,ylabel=$y$,
                ymin=-1,ymax=3,
                restrict y to domain=-1:4,
                enlargelimits={abs=0.5cm},
                axis line style={latex-latex},
                axis x line=center,
                axis y line=center,
                xtick={\empty},ytick={\empty},
                xlabel style={at={(ticklabel* cs:1)},anchor=west,font=\scriptsize},
                ylabel style={at={(ticklabel* cs:1)},anchor=north,font=\scriptsize},
                mark options={mark=*, mark size=0.6pt, gray, fill=gray},
            ]
            \addplot [thick,black,->] coordinates {(-\VIT,0) (0,0)} node[below left,pos=1,font=\normalsize] {$V_1$};
            \addplot table [x expr={sqrt((\VIIT*cos(\PhiT)+\SigT*\thisrow{x})*(\VIIT*cos(\PhiT)+\SigT*\thisrow{x})+(\VIIT*sin(\PhiT)+\SigT*\thisrow{y})*(\VIIT*sin(\PhiT)+\SigT*\thisrow{y}))*cos(\PhiT)}, y expr={sqrt((\VIIT*cos(\PhiT)+\SigT*\thisrow{x})*(\VIIT*cos(\PhiT)+\SigT*\thisrow{x})+(\VIIT*sin(\PhiT)+\SigT*\thisrow{y})*(\VIIT*sin(\PhiT)+\SigT*\thisrow{y}))*sin(\PhiT)}, only marks, mark=o] {points.dat};
            \addplot [thick,black,->] coordinates {(0,0) ({\VIIT*cos(\PhiT)},{\VIIT*sin(\PhiT)})} node[below right,pos=1,font=\normalsize] {$V_2$};
            \draw [ultra thin,black,->] (axis cs:1,0) arc[start angle=0, end angle=\PhiT, radius={transformdirectionx(1)}] node[left,pos=0.4,font=\normalsize] {$\phi$};
            \addplot [only marks, samples=1, black, mark size=0.5pt, line width=0.3pt, mark options={fill=gray}] ({\XT},{\YT});
            \addplot [thick,red,->] coordinates {(-\VIT,0) (\XT,\YT)} node[above,pos=0.5,font=\normalsize] {$\rho$};
            \addplot [thick,blue,->] coordinates {(0,0) (\XT,\YT)} node[left,pos=0.5,font=\normalsize] {$r$};
            \draw [ultra thin,red,->] (axis cs:-\VIT+2.0,0) arc[start angle=0, end angle={atan2(\YT,\VIT+\XT)}, radius={transformdirectionx(2.0)}] node[right,pos=0.6,font=\normalsize] {$\varphi$};
        \end{axis}
    \end{tikzpicture}
    };
    \caption{TWDP phasor diagram illustrating $p(r|\phi)$ for $\varPhi_1=0$}
    \label{Fig_2}    
\end{figure}
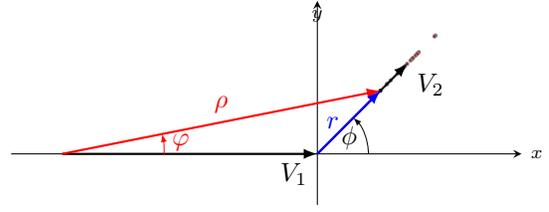

\noindent or alternatively, as~\cite[03.02.02.0001.01]{Wol}:
\begin{equation}
\label{prEq}
\begin{split}
    p(r)&=\sum_{m=0}^{\infty} \overbrace{\frac{\left(\frac{V_2^2}{2\sigma^2}\right)^{m} e^{-\frac{V_2^2}{2\sigma^2}}}{\Gamma[m+1]}}^{\mathclap{p(m):~m\sim{Po}\left(\frac{V_2^2}{2\sigma^2}\right)}}\\
    &\times \underbrace{\frac{2}{\Gamma[m+1]} r^{2m+1} \left(\frac{1}{2\sigma^2}\right)^{m+1} \exp{\left(-\frac{r^2}{2\sigma^2}\right)}}_{\mathclap{p(r|m):~r\sim\text{\textit{Nakagami}\big($m+1$,$2\sigma^2 (m+1)$\big)}}}
\end{split}
\end{equation} 

\noindent Considering the above, $p(\varphi|r)$ can be derived from $p(\phi|r)$ given in (\ref{Fi}) by applying a random variable transformation:
\begin{equation}
\begin{split}    \varphi=\arctan{\left(\frac{r \sin{\phi}}{V_1+r \cos{\phi}}\right)}
\end{split}
\end{equation} 
as: 
\begin{equation}
\label{p_fir}
\begin{split}
    p(\varphi|r)=
    \begin{cases}
      \frac{V_1 \cos{\varphi}}{\pi\sqrt{r^2-V_1^2 \sin^2\varphi}},&\text{$r<V_1$, $|\varphi|<\arcsin{\frac{r}{V_1}}$}\\
      \frac{1}{2\pi}+\frac{V_1 \cos{\varphi}}{2\pi\sqrt{r^2-V_1^2 \sin^2\varphi}},&\text{$r>V_1$, $\varphi\in\left(-\pi,\pi\right)$}\\
      0,&\text{otherwise}
    \end{cases} 
\end{split}
\end{equation} 
and the conditional PDF of the TWDP phase can be obtained after inserting (\ref{p_fir}) and (\ref{prEq}) into (\ref{ItemEq}), as:  
\begin{equation}
\label{eq9}
\begin{split}
    p(\varphi)&\Big|_{|\varphi|<\frac{\pi}{2}}=\sum_{m=0}^{\infty} \frac{\left(\frac{V_2^2}{2\sigma^2}\right)^{m} e^{-\frac{V_2^2}{2\sigma^2}}}{\Gamma[m+1]} \\
    &\times\left(
      \int\displaylimits_{V_1 |\sin{\varphi}|}^{V_1} p(r|m) \frac{1}{\pi}\frac{1}{\tan|\varphi|}\frac{V_1|\sin\varphi|}{\sqrt{r^2-V_1^2 \sin^2\varphi}}dr\right.\\
    &\left.+\int\displaylimits_{V_1}^{+\infty} p(r|m) \frac{1}{2\pi}\left(1+\frac{1}{\tan|\varphi|}\frac{V_1|\sin\varphi|}{\sqrt{r^2-V_1^2 \sin^2\varphi}}\right)dr\right)
\end{split}
\end{equation}

\noindent
\begin{equation}
\label{eq10}
\begin{split}
    p(\varphi)&\Big|_{|\varphi|>\frac{\pi}{2}}=\sum_{m=0}^{\infty} \frac{\left(\frac{V_2^2}{2\sigma^2}\right)^{m} e^{-\frac{V_2^2}{2\sigma^2}}}{\Gamma[m+1]} \\
    & \times \int\displaylimits_{V_1}^{+\infty} p(r|m) \frac{1}{2\pi}\left(1+\frac{1}{\tan|\varphi|}\frac{V_1|\sin\varphi|}{\sqrt{r^2-V_1^2 \sin^2\varphi}}\right)dr
\end{split}
\end{equation} 
\noindent Note that after some manipulations, both equations, (\ref{eq9}) and (\ref{eq10}), can be merged into the single expression and expressed as a sum of three different integrals:
\begin{equation}
\begin{split}
\label{PDF_preko_I}
    p(\varphi)&=\sum_{m=0}^{\infty} \overbrace{\frac{\left(\frac{V_2^2}{2\sigma^2}\right)^{m} e^{-\frac{V_2^2}{2\sigma^2}}}{\Gamma[m+1]}}^{\mathclap{p(m)}}\\
    &\times\underbrace{\left(\frac{1}{2\pi}\frac{1}{\tan{|\varphi|}}\mathcal{I}_{1}+\frac{1}{2\pi}\mathcal{I}_{2}+\frac{1}{2\pi}\frac{1}{|\tan{\varphi}|}\mathcal{I}_{x}\right)}_{\mathclap{p(\varphi|m)}}
\end{split}
\end{equation}
where these integrals are defined and solved respectively, as:
\begin{equation}
\begin{split}    \mathcal{I}_{1}&=\int\displaylimits_{V_1|\sin\varphi|}^{+\infty} \frac{p(r|m) V_1 |\sin{\varphi}|}{\sqrt{r^2-V_1^2 \sin^2\varphi}}dr=\Bigg|x=\frac{r^2}{2\sigma^2},~\text{\cite[2.3.6.6]{Pru86}}\Bigg|\\
    &=\frac{\sqrt{\pi}}{\Gamma[m+1]} \left(\frac{V_1^2 \sin^2\varphi}{2\sigma^2}\right)^{m+1}\exp\left(-\frac{V_1^2 \sin^2\varphi}{2\sigma^2}\right) \\
    &\times U\left(\frac{1}{2},m+\frac{3}{2},\frac{V_1^2 \sin^2\varphi}{2\sigma^2}\right)
\end{split}
\label{I1}
\end{equation} 
\begin{equation}
\begin{split}
    \mathcal{I}_{2}&=\int\displaylimits_{V_1}^{+\infty} p(r|m) dr=\Bigg|x=\frac{r^2}{2\sigma^2},~\text{\cite[2.3.6.6]{Pru86}}\Bigg|\\
    &=\frac{1}{\Gamma[m+1]}\left(\frac{V_1^2}{2\sigma^2}\right)^{m+1}\exp\left(-\frac{V_1^2}{2\sigma^2}\right) U\left(1,m+2,\frac{V_1^2}{2\sigma^2}\right)
\end{split}
\label{I2}
\end{equation} 
\setcounter{equation}{15}
\begin{equation}
\begin{split}
    \mathcal{I}_{x}&=\int\displaylimits_{V_1|\sin\varphi|}^{V_1} \frac{p(r|m) V_1 |\sin{\varphi}|}{\sqrt{r^2-V_1^2 \sin^2\varphi}}dr\\
    &=\Bigg|u=\frac{\frac{V_1^2-r^2}{2\sigma^2}}{\frac{V_1^2}{2\sigma^2 }\cos^2\varphi},~\text{\cite[p.348, eq. (9)]{Erd1940}}\Bigg|\\
    &=\frac{2}{\Gamma[m+1]} \frac{\sin^2\varphi}{|\tan{\varphi}|} \left(\frac{V_1^2}{2\sigma^2}\right)^{m+1} \exp{\left(-\frac{V_1^2}{2\sigma^2}\right)} \\
    & \times \Phi_1\left(1,-m,\frac{3}{2},\cos^2\varphi,\frac{V_1^2}{2\sigma^2}\cos^2\varphi\right)
\end{split}
\label{Ix}
\end{equation} 
\setcounter{equation}{16}
\begin{figure*}[ht]
\begin{equation}
  \begin{aligned}
   p(\varphi|\varPhi_1=0)&=\sum_{m=0}^{\infty} \frac{\left(\frac{V_2^2}{2\sigma^2}\right)^{m} e^{-\frac{V_2^2}{2\sigma^2}}}{\Gamma[m+1]}\left[\frac{1}{2\pi}
    \frac{1}{\Gamma[m+1]} \left(\frac{V_1^2}{2\sigma^2}\right)^{m+1}\exp\left(-\frac{V_1^2}{2\sigma^2}\right) U\left(1,m+2,\frac{V_1^2}{2\sigma^2}\right)\right.
   \\
   &\left.+\frac{1}{2\pi}\frac{\sqrt{\pi}}{\tan{|\varphi|}}\frac{1}{\Gamma[m+1]} \left(\frac{V_1^2 \sin^2\varphi}{2\sigma^2}\right)^{m+1}\exp\left(-\frac{V_1^2 \sin^2\varphi}{2\sigma^2}\right) U\left(\frac{1}{2},m+\frac{3}{2},\frac{V_1^2 \sin^2\varphi}{2\sigma^2}\right)\right.\\
   &\left.+\frac{1}{2\pi} 2\cos^2{\varphi}\frac{1}{\Gamma[m+1]}  \left(\frac{V_1^2}{2\sigma^2}\right)^{m+1} \exp{\left(-\frac{V_1^2}{2\sigma^2}\right)} \Phi_1\left(1,-m,\frac{3}{2},\cos^2\varphi,\frac{V_1^2 \cos^2\varphi}{2\sigma^2}\right) \right]
  \end{aligned}
  \label{faza}
\end{equation}
\begin{equation}
\label{faza_gen}
  \begin{aligned}
   p(\varphi|\varPhi_1)&=\sum_{m=0}^{\infty} \frac{\left(\frac{V_2^2}{2\sigma^2}\right)^{m} e^{-\frac{V_2^2}{2\sigma^2}}}{\Gamma[m+1]}\left[\frac{1}{2\pi}
    \frac{1}{\Gamma[m+1]} \left(\frac{V_1^2}{2\sigma^2}\right)^{m+1}\exp\left(-\frac{V_1^2}{2\sigma^2}\right) U\left(1,m+2,\frac{V_1^2}{2\sigma^2}\right)\right.\\
    &\left.+\frac{1}{2\pi}\frac{\sqrt{\pi}}{\tan{|\varphi-\varPhi_1|}}\frac{1}{\Gamma[m+1]} \left(\frac{V_1^2 \sin^2(\varphi-\varPhi_1)}{2\sigma^2}\right)^{m+1}\exp\left(-\frac{V_1^2 \sin^2(\varphi-\varPhi_1)}{2\sigma^2}\right) U\left(\frac{1}{2},m+\frac{3}{2},\frac{V_1^2 \sin^2(\varphi-\varPhi_1)}{2\sigma^2}\right)\right.\\
    &\left. +\frac{1}{2\pi} 2\cos^2{(\varphi-\varPhi_1)}\frac{1}{\Gamma[m+1]}  \left(\frac{V_1^2}{2\sigma^2}\right)^{m+1} \exp{\left(-\frac{V_1^2}{2\sigma^2}\right)} \Phi_1\left(1,-m,\frac{3}{2},\cos^2(\varphi-\varPhi_1),\frac{V_1^2 \cos^2(\varphi-\varPhi_1)}{2\sigma^2}\right)\right]
  \end{aligned}
\end{equation}
\end{figure*}

As a result, the final expression for the conditional phase PDF of the TWDP multipath fading process is given by (\ref{faza}), where $U(\cdot,\cdot,\cdot)$ is the Tricomi confluent hypergeometric function and $\Phi_1(\cdot,\cdot,\cdot;\cdot,\cdot)$ is the Humbert confluent hypergeometric function 
(both of which can be easily evaluated and efficiently implemented in most standard software packages, such as Matlab and Mathematica), while the general expression for the arbitrary phase of the stronger specular component is given by (\ref{faza_gen}).

Since TWDP model includes the Rice ($V_2=0$) and Rayleigh ($V_1=0$) models as special cases, the correctness of (\ref{faza_gen}) can be confirmed by its reduction to these well-known phase PDFs. 
Specifically, for $V_2=0$, only the first summation term (i.e. $m=0$) exists in (\ref{faza_gen}), so according to \cite[07.33.03.0005.01]{Wol}, and \cite[4.18]{Bry2012}, \cite[07.20.17.0013.01]{Wol}, \cite[06.25.26.0001.01]{Wol}, equation (\ref{faza_gen}) reduces to the well-known form of the Rician phase PDF~\cite[eq. (45)]{beckmann1964rayleigh}. The conditional probability density function of the Rician PDF further reduces to $1/(2\pi)$ when the dominant specular component vanishes ($V_1=0$), as expected for Rayleigh fading.

\section{Closed-form expression}
\label{sec:III}
In this section, the closed-form expression for the conditional TWDP phase PDF is derived, starting from its infinite series representation given by (\ref{faza_gen}). To obtain it's closed-form counterpart, expression (\ref{faza_gen}) is treated as the sum of three separated series.

To solve the first series, \cite[07.33.03.0014.01]{Wol} and \cite[06.06.07.0001.01]{Wol} are used, and the order of integration and summation is reversed. The resulting sum and integral are then solved with the help of \cite[03.02.02.0001.01]{Wol}, and \cite[eq. (3.12)]{Bry2012}, respectively, and expressed in terms of the Humbert confluent hypergeometric function $\Phi_3(\cdot,\cdot,\cdot)$.

The second series is solved with the help of \cite[07.33.17.0007.01]{Wol}, \cite[07.33.03.0052.01]{Wol}, and \cite[eq. (5.11.4.11)]{Pru2}, and is also expressed in terms of the $\Phi_3(\cdot,\cdot,\cdot)$ function.

The third series is then solved using~\cite[Sec. 1.6 eq. (36)]{Srivastava1984} and \cite[Sec. 1.7 eq. (39)]{Srivastava1984}, and the final closed-form expression of the conditional TWDP phase is given by (\ref{zatvorena}), where $F^{(3)}\left[\cdot,\cdot,\cdot\right]$ is the general triple hypergeometric function.

\begin{figure*}[ht]
\begin{equation}
\label{zatvorena}
  \begin{aligned}
   p(\varphi|\varPhi_1)&=\frac{1}{2\pi}\left[1-\frac{V_1^2}{2\sigma^2}e^{-\frac{V_1^2+V_2^2}{2\sigma^2}}\Phi_3\left(1,2,\frac{V_1^2}{2\sigma^2},\left(\frac{V_1 V_2}{2\sigma^2}\right)^2\right)\right]\\
    &+\frac{\sqrt{\pi} }{2\pi} \sqrt{\frac{V_1^2}{2\sigma^2}} \cos\left(\varphi-\varPhi_1\right) e^{-\frac{V_1^2\sin^2\left(\varphi-\varPhi_1\right)+V_2^2}{2\sigma^2}}\Phi_3\left(\frac{1}{2},1,\frac{V_2^2}{2\sigma^2},\left(\frac{V_1 V_2 \sin\left(\varphi-\varPhi_1\right)}{2\sigma^2}\right)^2\right)\\
    &+\frac{2}{2\pi} \frac{V_1^2}{2\sigma^2} \cos^2{\left(\varphi-\varPhi_1\right)} e^{-\frac{V_1^2+V_2^2}{2\sigma^2}}\\
    &\times 
    F^{(3)}\left[
\begin{array}{c}
()::(1);\\
()::(\frac{3}{2});
\end{array}
\begin{array}{c}
();\\
();
\end{array}
\begin{array}{c}
(\phantom{1}):();\\
(1):();\\
\end{array}
\begin{array}{c}
();\\
();\\
\end{array}
\begin{array}{c}
();\\
();\\
\end{array}
-\frac{V_1^2}{2\sigma^2}\frac{V_2^2}{2\sigma^2} \cos^2{\left(\varphi-\varPhi_1\right)},\frac{V_1^2}{2\sigma^2} \cos^2{\left(\varphi-\varPhi_1\right)}, \frac{V_1^2}{2\sigma^2}\frac{V_2^2}{2\sigma^2}
\right]
  \end{aligned}%
\end{equation}
\end{figure*}

Accordingly, the closed form of the conditional TWDP phase PDF involves Humbert confluent and general triple hypergeometric functions, which are less known and relatively complex. Therefore, it is necessary to investigate the convergence of (\ref{zatvorena}), which can be easily proven with the help of \cite[p. 28]{Ribeiro}:
\setcounter{equation}{19}
\begin{equation}
\begin{split}
    \label{konvergencija}
    \left|\Phi_3(b,c;w,z)\right|\leq e^{2\sqrt{|z|}+2|w|+\frac{|b w|}{|c|}}
\end{split}
\end{equation} 
and \cite[Sec. 1.7 eq. (41)]{Srivastava1984}.

Despite that, 
its practical implementation and interpretation of its results are complex. Therefore, in the following section, a truncation analysis of the infinite series representation of the conditional TWDP phase PDF, given in (\ref{faza}) or (\ref{faza_gen}), is provided and the obtained expression is used for all subsequent  analysis.

\section{Truncation Error of the Infinite-series Expression}
\label{sec:IV}

The truncated sum, as a  method, reduces the complexity of the analysis and data processing, making them more appropriate for practical application. This approach not only simplifies the calculations, but also enhances the model's robustness, allowing better interpretation of the results. Below, the truncation error analysis is presented, followed by concrete examples to illustrate its application.

Since the conditional PDF of the TWDP phase (\ref{faza}) is expressed as an infinite-series, it is essential to determine the finite number of summation terms to achieve the  desired truncation error. Therefore, let's begin with the initial integral expression (\ref{ItemEq}):
\setcounter{equation}{20}
\begin{equation}
    \begin{split}
        p(\varphi)=\int p(\varphi|r) p(r) dr
    \end{split}
\end{equation}
where $p(r)$, according to (\ref{prEq}), can be expressed in series form as:
\begin{equation}
    \begin{split}
        p(r)=\sum_m p(m) p(r|m) 
    \end{split}
\end{equation} 
\noindent where the conditional random variable $r|m$ follows a Nakagami-\textit{m} distribution, i.e. \newline
$r\sim$~\textit{Nakagami}-\textit{m}\big($m+1$,$2\sigma^2 (m+1)$\big), while $m$ is a Poisson distributed random variable, i.e. $m\sim \textit{Po}\left(V_2^2/(2\sigma^2)\right)$.
Accordingly, the conditional PDF of the TWDP phase is obtained as the following series:
\begin{equation}
    \begin{split}
        p(\varphi)&=\sum_m p(m) \underbrace{\int p(\varphi|r) p(r|m) dr}_{\mathclap{p(\varphi|m)}}
    \end{split}
\end{equation}

\noindent where $p(\varphi|m)$ is determined in its exact form, as given (see (\ref{faza})). Consequently, the truncation error of the conditional phase PDF will be directly dependent on the truncation error of $p(r)$. So, to assess the truncation error of the distribution $p(r)$, let's adopt the measure that $\alpha$\% of the total average power of the variable $r$ is preserved in the truncated sum. To satisfy this requirement, the total average power of the variable $r$ can be expressed as:
\begin{equation}
\begin{split}
    \overline{r^2}=V_2^2+2\sigma^2 &=\sum_{m=0}^{\infty} p(m) 2\sigma^2 (m+1)
\end{split}
\end{equation} 
which, after the total average power normalization, allows us to calculate the share of the total average power over the particular summation terms as:
\begin{equation}
\begin{split}
    1 &=\sum_{m=0}^{\infty} \underbrace{\frac{\left(\frac{V_2^2}{2\sigma^2}\right)^{m} e^{-\frac{V_2^2}{2\sigma^2}}}{\Gamma[m+1]} \frac{2\sigma^2 (m+1)}{V_2^2+2\sigma^2}}_{p_P(m)}
\end{split}
\end{equation} 
Since $p_P(m)$ is obviously normalized, it represents PDF of the average power over the summation terms, given as:
\begin{equation}
\begin{split}
    p_p(m) &=\frac{\left(\frac{V_2^2}{2\sigma^2}\right)^{m} e^{-\frac{V_2^2}{2\sigma^2}}}{\frac{\Gamma[m+1]}{m+1}} \frac{1}{1+\frac{V_2^2}{2\sigma^2}},\text{   for }m\geq0
\end{split}
\end{equation} 
which can be written as:
\begin{equation}
\begin{split}
    p_p(m) &= \frac{\nu}{1+\nu} \text{Po}\left(m-1,\nu\right)\\
    &+ \frac{1}{1+\nu} \text{Po}\left(m,\nu\right), \text{~~for~} m\geq 0
\end{split}
\end{equation} 
where $\nu=V_2^2/(2\sigma^2)$ and $\text{Po}(k,\nu)=\nu^k/k! \exp(-\nu)$.
Clearly, the amount of average power across the summation terms follows a weighted Poisson distribution. Thus, the boundaries of the summation terms, which preserve $\alpha$\% of the total average power of the variable $r$, can be determined as the $\alpha$\% confidence interval of the average power distribution over these summation terms.

\begin{figure}[b]
    \vspace{-0.4cm}
	\centering
    \includegraphics[width=.425\textwidth]{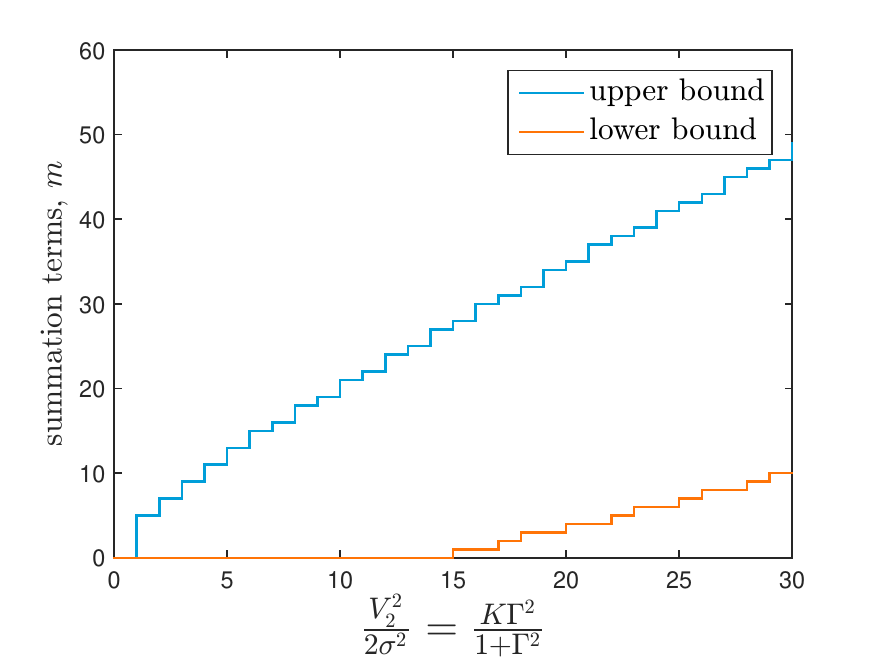}
	\caption{Boundaries of the parameter $m$ which achieve defined truncation error of the conditional TWDP phase PDF}
	\label{FigTrankiranje}
\end{figure}

Considering the aforesaid, the confidence interval of the weighted Poisson distribution can be determined as the subset of the union of the confidence intervals of each individual Poisson variable:
\begin{equation}
    \begin{split}
        \text{CI}_{\alpha\%}  &\subseteq \text{CI}_{\alpha\%}\left[\text{Po}\left(m-1,\nu\right)\right] \cup \text{CI}_{\alpha\%}\left[\text{Po}\left(m,\nu\right)\right]
    \end{split}
\end{equation}
where Wald's confidence interval~\cite{Pat2012} for each variable can be obtained  as:
\begin{equation}
    \begin{split}
        \text{CI}&_{\alpha\%}\left[\text{Po}\left(m-1,\nu\right)\right] \\
        &:= \Set{m-1}{0\leq \nu+Z_{\alpha_1}\sqrt{\nu}\leq m \leq \nu+Z_{\alpha_2}\sqrt{\nu}}\\
        \text{CI}&_{\alpha\%}\left[\text{Po}\left(m,\nu\right)\right] \\
        &:= \Set{m}{0\leq \nu+Z_{\alpha_1}\sqrt{\nu}\leq m \leq \nu+Z_{\alpha_2}\sqrt{\nu}}
    \end{split}
\end{equation}
and
\begin{equation}
    \begin{split}
        \alpha_1&=\frac{1}{2}-\frac{\alpha\%}{200} ~~~~~~~~
        \alpha_2=\frac{1}{2}+\frac{\alpha\%}{200}\\
        Z_a&=\Bigg\{y\Bigg|\int_{-\infty}^{y} \frac{1}{\sqrt{2\pi}} \exp{\left(-\frac{x^2}{2}\right)}\diff{x}=a\Bigg\}
    \end{split}
\end{equation}

Accordingly, the boundaries of the truncated sum, which preserve at least $\alpha\%$ of the total average power of random variable $r$, can be determined as:
\begin{equation}
    \begin{split}
        \text{CI}_{\alpha\%}  &\subseteq \SET{m}{$0\leq \nu-1+Z_{\alpha_1}\sqrt{\nu}\leq m \leq \nu+Z_{\alpha_2}\sqrt{\nu}$}
    \end{split}
\end{equation}
i.e. \begin{equation}
\begin{split}
    0 \leq \nu-1+Z_{\alpha_1}\sqrt{\nu}\leq m \leq \nu+Z_{\alpha_2}\sqrt{\nu}
\end{split}
\end{equation} 

Considering the aforementioned, the truncated conditional TWDP phase PDF, which preserves 99.9\% of the total average power, can be calculated using the following expression:
\begin{equation}
\begin{split}
    p(\varphi)&=\sum_{m=\floor[\bigg]{\max\left\{0,\frac{V_2^2}{2\sigma^2}-1-3.291\sqrt{\frac{V_2^2}{2\sigma^2}}\right\}}}^{\ceil[\bigg]{\frac{V_2^2}{2\sigma^2}+3.291\sqrt{\frac{V_2^2}{2\sigma^2}}}} p(m) p(\varphi|m)
\end{split}
\label{granice}
\end{equation} 

Fig.~\ref{FigTrankiranje} shows the upper and the lower bounds of the summation terms given by (\ref{granice}), which must be considered to achieve defined truncation error in terms of $\frac{V_2^2}{2\sigma^2}$. These bounds are also expressed in terms of the TWDP model's parameters $K$, $\Gamma$, and $\Omega$, defined as~\cite{rad}: $K=(V_1^2+V_2^2)/2\sigma^2$, $\Gamma=V_2/V_1$ for $0\leq V_2 \leq V_1$, and $\Omega=V_1^2+V_2^2+2\sigma^2$, as these parameters can be more easily correlated with different channel conditions.

As it can be observed from Fig.~\ref{FigTrankiranje}, the derived phase PDF expression converges very quickly, especially for small values of $K$ and $\Gamma$. Also, even for very large values of these parameters (e.g. when $K=60$ and $\Gamma=1$) the number of summation terms required to preserve 99.9\% of the total average power remains  relatively low (only 40, ranging from 10 to 49). Accordingly, the summation terms determined in (\ref{granice}) are used for all calculations of the conditional PDF of the TWDP phase, $p(\varphi)$, in the following sections.  

\begin{figure*}[t]
  \vspace{-0.4cm}
  \centering
  \subfloat[\label{Fig1}]{\includegraphics[width=.475\textwidth]{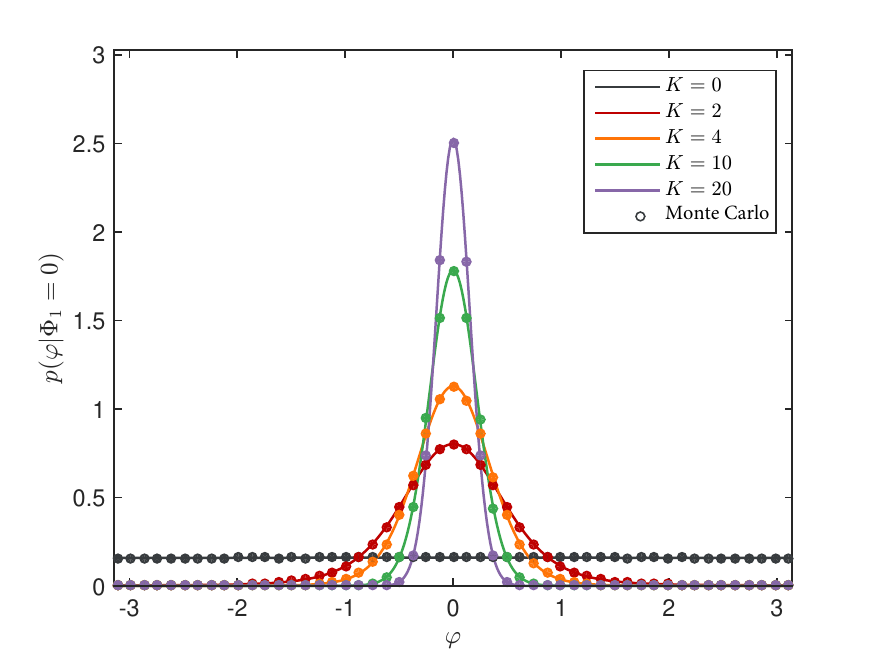}}\quad
  \subfloat[\label{Fig2}]{\includegraphics[width=.475\textwidth]{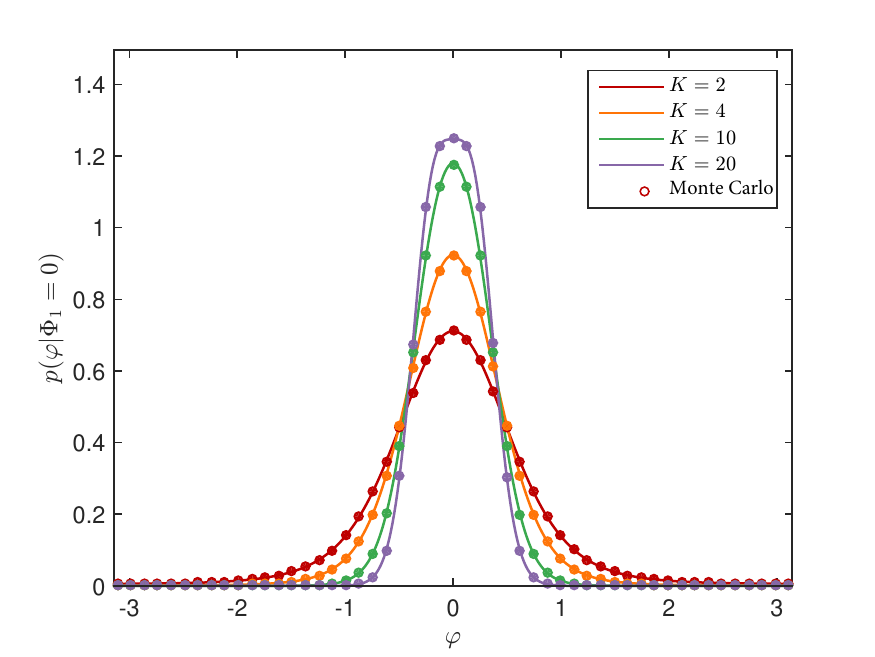}}\\
  \subfloat[\label{Fig3}]{\includegraphics[width=.475\textwidth]{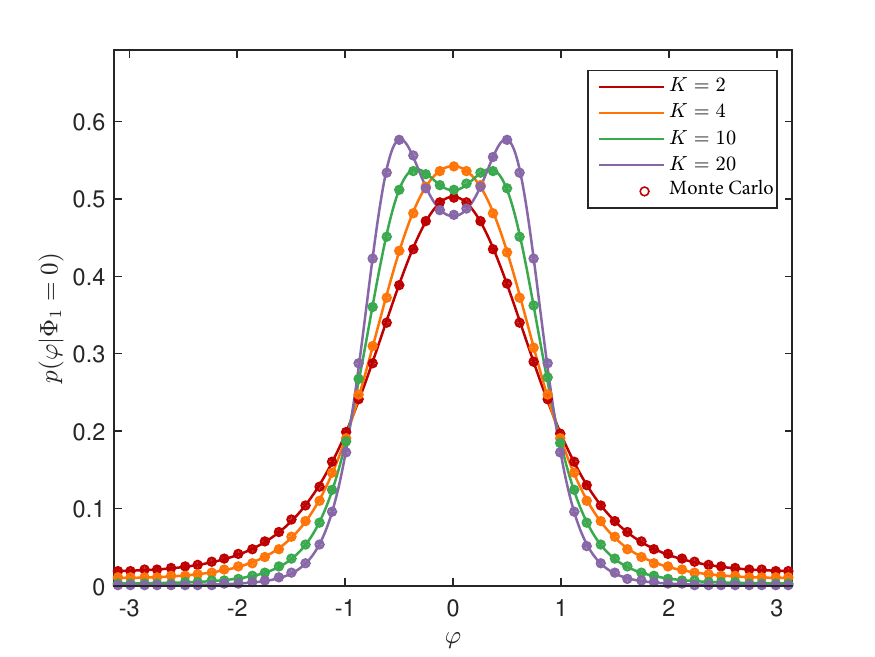}}\quad
  \subfloat[\label{Fig4}]{\includegraphics[width=.475\textwidth]{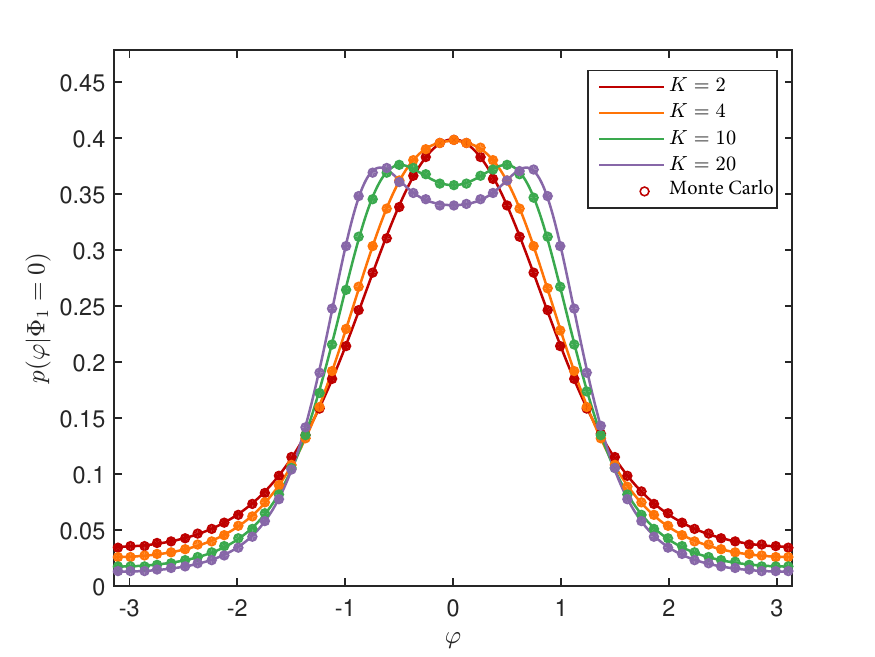}}
  \caption{The conditional TWDP phase PDF for (a) $\Gamma=0$ (b) $\Gamma=0.3$ (c) $\Gamma=0.7$ (d) $\Gamma=1$}
\end{figure*}

\section{Numerical results}
\label{sec:V} 

In this section, obtained analytical results are plotted to demonstrate the influence of different channel conditions on the behavior of the conditional TWDP phase. The validity of the derived expression (\ref{faza}) is also examined, by comparing theoretical curves against simulated results, which are generated using Monte Carlo simulation and $10^7$ samples. 
As shown in Fig.~\ref{Fig1} - Fig.~\ref{Fig4}, there is an excellent match  between simulation and analytical results for all considered cases.

In the initial step, the conditional TWDP phase PDF for Rician fading channel conditions is plotted in Fig.~\ref{Fig1} by setting $\Gamma=0$ (which occurs when only one specular component is present) and varying the values of $K$, in order to provide the reference point for further examination. The figure shows that for $K=0$ (i.e. when both specular components are absent and only diffuse component exists), the conditional phase follows an uniform distribution, as expected in Rayleigh channel. For other values of $K>0$, the obtained PDFs exhibit a symmetrical Gaussian-like shape, typical of Rician fading. In these cases, the phase variance decreases as $K$ increases, and the distribution can be well approximated by the Tikhonov (von Mises) PDF.

Figures~\ref{Fig2}–\ref{Fig4} illustrate other cases typical for channels with TWDP fading. These figures demonstrate that the appearance of the second specular component changes the phase PDF shape, simultaneously increasing its variance compared to the Rician case. These figures also show that the observed variance's increment becomes more pronounced for larger values of $\Gamma$, i.e., as  the magnitude of the second specular component increases.
More specifically, for small values of $\Gamma$ or small values of $K$, which describe better-than-Rayleigh fading conditions~\cite{rad} (where the observed fading is less severe than in Rayleigh channels~\cite{rad}), the conditional TWDP phase PDF exhibits a Gaussian-like unimodal shape with even symmetry. This behavior is similar to that observed in Rician fading channels, suggesting that any topic related to the phase estimation can be treated in a similar manner to Rician channels. 

However, as both $\Gamma$ and $K$ increase (i.e. when the power of two specular components are similar and significantly larger than the power of the diffuse component), the phase PDF becomes bimodal. 
It is shown in~\cite{rad} that such combination of parameters corresponds to a very severe fading, which is classified as  worse-than-Rayleigh. Accordingly, in channels with worse-than-Rayleigh fading, the phase PDF is spread out and exhibits bimodal behavior. In such cases, it is necessary to examine the phase error behavior obtained within the synchronization procedure, which is going to be significantly different from that observed in Rician channels.

\section{Verification of the Derived Analytical Expressions}

\label{sec:VI}

In this section, we verified theoretically obtained results using developed geometrically-based simulator which reflects propagation conditions in TWDP channel. To achieve this, propagation environment is constructed to closely align with the theoretical assumptions. It is assumed that the stronger specular component arrives to the receiver with constant envelope $V_1$ and constant phase $\phi_1$, while the weaker specular component propagates to the receiver with constant envelope $V_2$ and random phase $\phi_2$. It is also assumed that the diffuse component $n(t)$ results from the isotropic 2-D scattering, as Clarke suggested in~\cite{Cla68}.


\begin{figure*}[b]
  \vspace{-0.4cm}
  \centering
  \includegraphics[width=.675\textwidth]{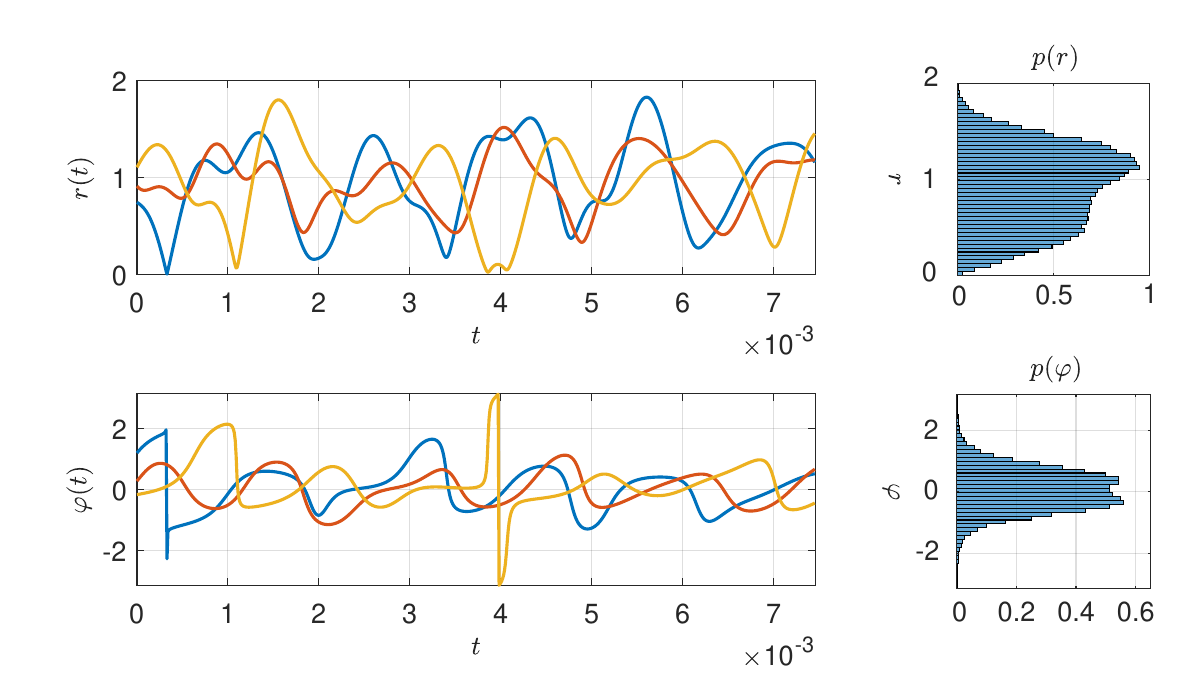}
  \caption{Three envelope and phase realizations of the TWDP fading process and corresponding histograms (for $K=10$, $\Gamma=0.7$, $\Omega=1$, $T_s\cdot f_d=0.01$)}
  \label{Figure5}
\end{figure*}

\begin{figure}[t]
    \vspace{-0.4cm}
	\centering
        \includegraphics[width=.425\textwidth]{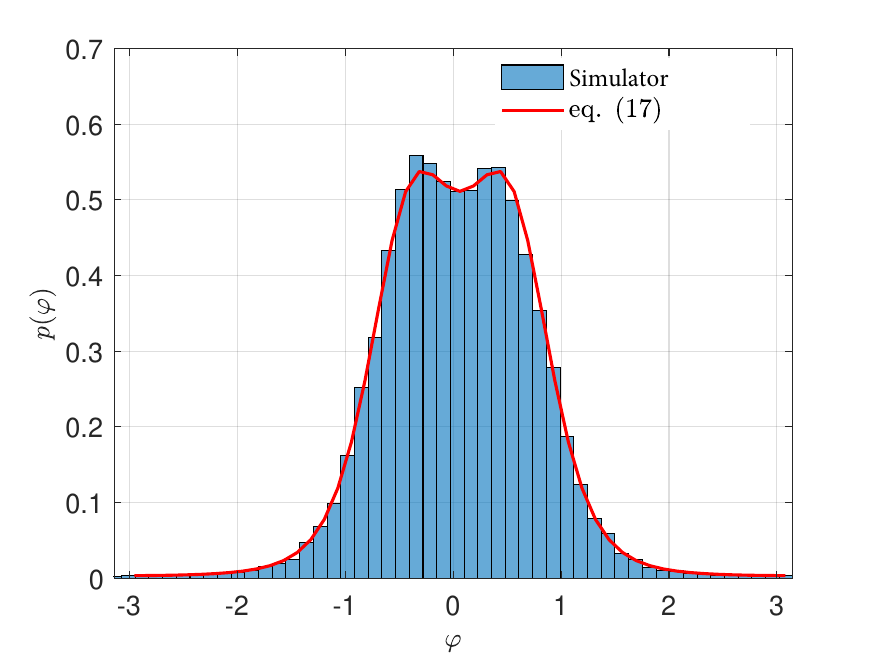}
	\caption{Comparison of analytical conditional TWDP phase PDF with corresponding histogram (for $K=10$, $\Gamma=0.7$, $\Omega=1$, $T_s\cdot f_d=0.01$)}
	\label{simulator}
\end{figure}

The propagation environment supporting these theoretical assumptions is constructed as below. 
The transmitter is fixed at the coordinates (0, 5~m), while the receiver, positioned at the origin of the global coordinate system, moves with a constant velocity $v$=10 m/s in the direction of the x-axis. The reflective plane is placed in the receiver's vicinity at coordinates (3.6633~m, -4.0613~m), at an angle of 25.0244${}^\circ$. 
When signal is emitted, the stronger specular component propagates along the line-of-sight (LoS) between the transmitter and the receiver and arrives perpendicular to the direction of motion. As such, a total phase of the stronger specular component is almost unaffected by the Doppler effect. On the other hand, the azimuthal angle of arrival of the weaker specular component, determined by relative positions of the receiver, transmitter and reflector, causes its total phase to change much faster than the phase of the stronger specular component.
Considering the aforesaid, simulation time is set to 7.5$\cdot$10${}^{-3}$s, in which LOS component rotates for  10${}^\circ$, while the weaker specular component makes 5 full rotations.
In addition, the maximum Doppler frequency is set to 1000 Hz, which corresponds to the carrier frequency of 30 GHz for a chosen velocity. The sample time is chosen in way that $T_s\cdot f_d=0.01$, i.e., sample time is set to $10^{-5}$s. On the other side, diffuse component is simulated as described in~\cite{Xia06}. To average the effect of diffuse component on the overall phase distribution, $200$ different realizations are taken into consideration, and used to obtain histograms given in Fig.~\ref{Figure5} and Fig.~\ref{simulator}.

Three different realizations of the amplitude and the phase of the TWDP process are obtained using described simulator and illustrated in Fig.~\ref{Figure5}, together with corresponding histograms which exhibit expected bimodal behavior. In Fig.~\ref{simulator}, TWDP phase histogram obtained by the simulator is compared to the derived analytical phase PDF (\ref{faza}), showing an excellent match between them, thus providing additional verification of the derived expressions.

\section{Performance Analysis}
\label{sec:VII}
\begin{figure*}[t]
  \vspace{-0.4cm}
  \centering
  \subfloat[\label{Fig11}]{\includegraphics[width=.475\textwidth]{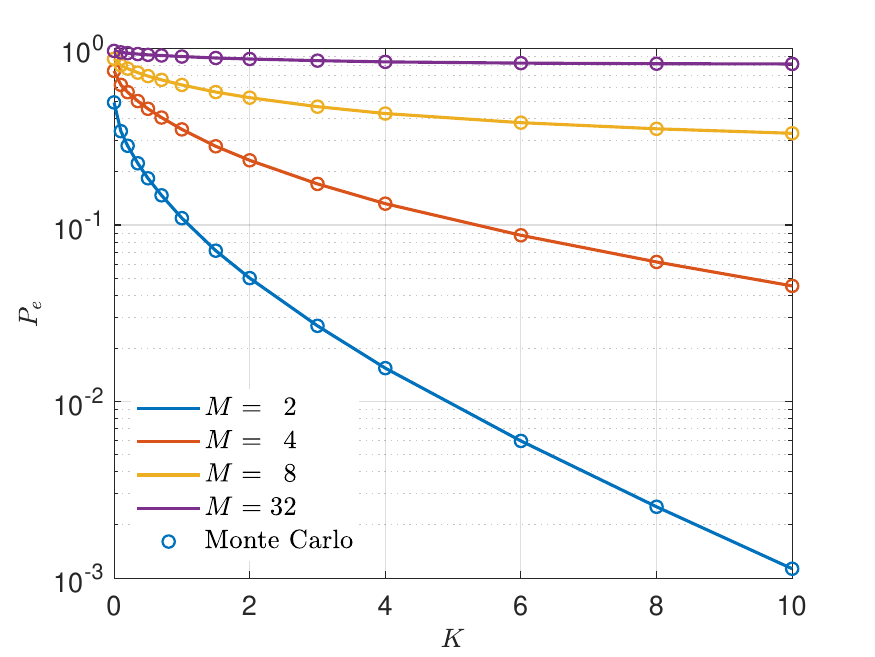}}\quad
  \subfloat[\label{Fig22}]{\includegraphics[width=.475\textwidth]{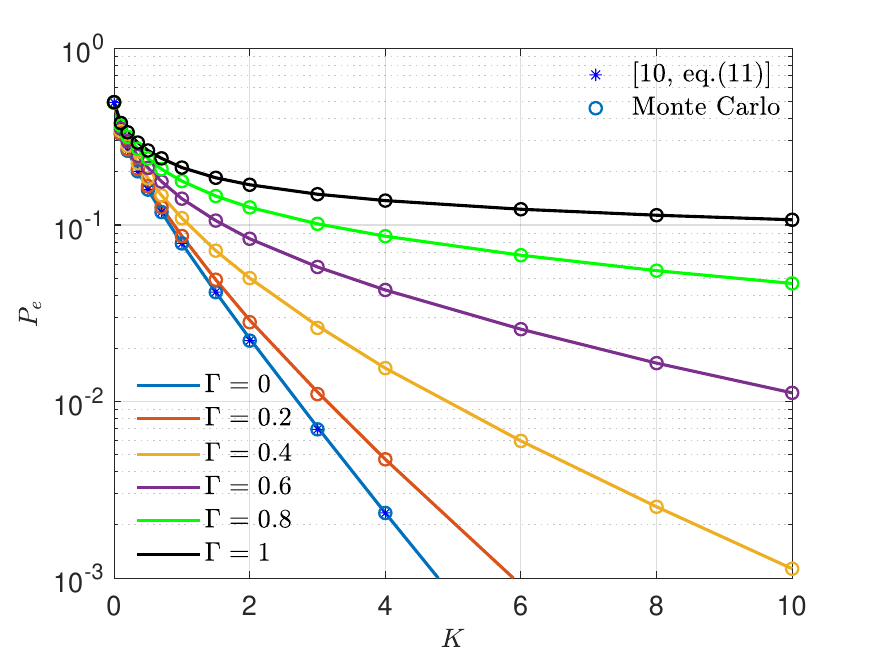}}
  \caption{$P_e$ in TWDP channel, given in terms of $K$ and (a) $\Gamma=0.4$ for $M$-ary PSK modulation (b) different values of $\Gamma$ for BPSK modulation}
\end{figure*}

To demonstrate the applicability of the derived phase PDF, the error probability ($P_e$), introduced in~\cite{Bro19}, is determined for the $M$-ary Phase Shift Keying (PSK) modulation scheme. Considered error probability, $P_e$, caused by imperfect phase synchronization without considering  impact of the white Gaussian noise, is defined as~\cite{Bro19}:

\begin{equation}
P_{e}=2 \int_{\pi/M}^{\pi} p(\varphi) \,d\varphi
\label{Pe}
\end{equation}
where $M$ is the modulation order and $p(\varphi)$ is the phase PDF. Considering the aforesaid, $P_e$ for TWDP channel can be obtained by inserting (\ref{faza}) into (\ref{Pe}).
Unfortunately, due to the complexity of the integral, given expression is evaluated numerically and the results are presented in Fig.~\ref{Fig11} - Fig.~\ref{Fig22}, together with those obtained using Monte Carlo simulation, which are in perfect accordance with analytical results.

The results indicate that for higher modulation orders the parameter $K$ has a negligible impact on $P_e$, where $P_e$  is high regardless of the value of $K$ (i.e. the severity of fading). In contrast, for BPSK modulation, as the parameter $K$ increases (i.e. as fading severity decreases), the error probability decreases significantly. 

A similar conclusion can be drawn from Fig.~\ref{Fig22}, obtained for BPSK modulation scheme exclusively. Obviously, in channels in which multipath fading is not particularly severe ($\Gamma \leq 0.6$), $P_e$ changes as a function of channel conditions (i.e. specific values of $K$ and $\Gamma$). In contrast, for channels with the pronounced multipath fading ($\Gamma \geq 0.8$), the values of $P_e$ are very high and are minimally influenced by the change of $K$. 

The Fig.~\ref{Fig22} is additionally important since one of the curves, obtained for $\Gamma=0$, can be compared with the results provided in~\cite{Bro19}. 
To facilitate this comparison, the limiting expression of $P_e$ derived for a composite Rician-Rician channel in~\cite[eq. (11)] {Bro19} is plotted for no-shadowing case (for $m \to \infty$). This curve (represented by asterisks) illustrates the behavior of $P_e$ in Rician channels and clearly matches the results obtained by combining equations (\ref{Pe}) and (\ref{faza}) for $\Gamma = 0$, which once again validates the accuracy of the derived analytical expressions.

\section{Conclusion}
In this paper, the exact closed-form and the infinite-series expressions for the conditional TWDP phase PDF are derived and used to investigate the impact of various propagation conditions on the behavior of the corresponding phase process. It is shown that, under better-than-Rayleigh fading conditions, the conditional phase PDF is unimodal, similar to the behavior observed in Rician channels. In such cases, the performance of modulation systems with non-ideal coherent detection can be assessed using already known principles. On the contrary, with the deterioration of fading conditions, the phase PDF becomes bimodal and spreads over the wider range of phase values $\varphi$. Therefore, under these conditions, all aspects related to phase estimation, need to be reconsidered.

\section*{Acknowledgment}
The authors would like to thank Prof. Ivo Kostić for many valuable discussions and advice. Also, the authors are greatly thankful to Prof. Nikolay Savischenko for his help in obtaining closed-form expressions.

\bibliographystyle{IEEEtran}

	





\end{document}